\documentclass[aip,jcp]{revtex4-1}
\usepackage{xr-hyper} 

\usepackage{amsmath}
\usepackage{rotating}
\usepackage{graphicx}
\usepackage{hyperref}
\usepackage{epstopdf}
\usepackage{color}
\usepackage{booktabs}
\usepackage{longtable}
\usepackage{float}

\usepackage{color}
\usepackage{colortbl}
\usepackage{natbib}
\usepackage[version=3]{mhchem} 

\usepackage{tabularx} 
\usepackage[table,xcdraw,dvipsnames]{xcolor}
\usepackage{braket} 
\usepackage{here} 
\usepackage[section]{placeins} 
\usepackage{siunitx} 
\usepackage{pifont} 
\usepackage[nolist]{acronym} 
\usepackage[normalem]{ulem}
\definecolor{comm}{rgb}{0,0,0}
\definecolor{comm2}{rgb}{0,0,0}
\definecolor{comm3}{rgb}{0,0,0}

\begin{document}

\begin{acronym}
        \acro{OSC}{open-shell character}
        \acro{CSC}{closed-shell character}
        \acro{OS}{open-shell state}
        \acro{CS}{closed-shell state}
        \acro{BLA}{bond-length alternation}
        \acro{VT}{variable temperature}
        \acro{LST}{linear synchronous transit}
        \acro{NLO}{nonlinear optical}
        \acro{DTE}{Dithienylethene}
        \acro{ESR}{electron spin resonance}
        \acro{PES}{potential energy surface}
        \acro{KS}{Kohn--Sham}
        \acro{DFT}{density functional theory}
        \acro{xc}{exchange-correlation}
\end{acronym}


\title{\Large Structural diradical character}


\author{B. Alexander Voigt}
\affiliation
{Institute for Inorganic and Applied Chemistry, Martin-Luther-King-Platz 6, University of Hamburg, 20146 Hamburg, Germany}
\author{Torben Steenbock}
\affiliation
{Institute of Physical Chemistry, Grindelallee 117, University of Hamburg, 20146 Hamburg, Germany}
\author{Carmen Herrmann}
\email{Carmen.Herrmann@chemie.uni-hamburg.de}
\affiliation
{Institute for Inorganic and Applied Chemistry, Martin-Luther-King-Platz 6, University of Hamburg, 20146 Hamburg, Germany}

\date{\today}

\begin{abstract}
A reliable first-principles description of singlet diradical character is essential for predicting nonlinear optical and magnetic properties of molecules. Since diradical and closed-shell electronic structures differ in their distribution of single, double, triple and aromatic bonds, modeling electronic diradical character requires accurate bond-length patterns, in addition to accurate absolute bond lengths. We therefore introduce structural diradical character, which we suggest as an additional measure for comparing first-principles calculations with experimental data.
We employ this measure to identify suitable exchange--correlation functionals for predicting the bond length patterns and electronic diradical character of a biscobaltocene with the potential for photoswitchable nonlinear optical activity. Out of four popular approximate exchange--correlation functionals with different exact-exchange admixtures (BP86, TPSS, B3LYP, TPSSh), the two hybrid functionals TPSSh and B3LYP perform best for diradical bond length patterns, with TPSSh being best for the organometallic validation systems and B3LYP for the organic ones. Still, none of the functionals is suitable for correctly describing relative bond lengths across the range of molecules studied, so that none can be recommended for predictive studies of (potential) diradicals without reservation. 
\end{abstract}


\maketitle



\section{Introduction}\label{intro}
\label{sec:intro}


Open-shell singlet diradical molecules have aroused interest among both theoreticians and experimentalists due to their special physical and chemical properties.\cite{intro1,intro2,intro3,intro4,jung03,mich96,sale72,line11,lenh10,stuy17} Among those are their \ac{NLO} properties, especially the second  \textcolor{comm2}{hyperpolarizability}, which can be tuned and amplified by a change in open-shell diradical character $y$~\cite{naka13,nand17,nlo1,nlo2,nlo4,nlo5,naka16}. 
Nonlinear optically active molecular materials are
important for applications such as 
data storage and telecommunication~\cite{muha16}.

It would be particularly useful if the \ac{NLO} properties of molecules or molecular materials could be switched by external stimuli. There has been considerable experimental\cite{nlo-switch-exp1,nlo-switch-exp2,sun15,atta17,cari17} and theoretical\cite{schu15,nlo-switch-exp2,nlo3,torr17,flor17} work on switchable organic and organometallic \ac{NLO}-active molecules, showing that a variety of stimuli such as pH, temperature, redox reactions, and light can be used for this purpose.

\begin{figure}[ht]
	\centering
		\includegraphics[width=0.666\textwidth]{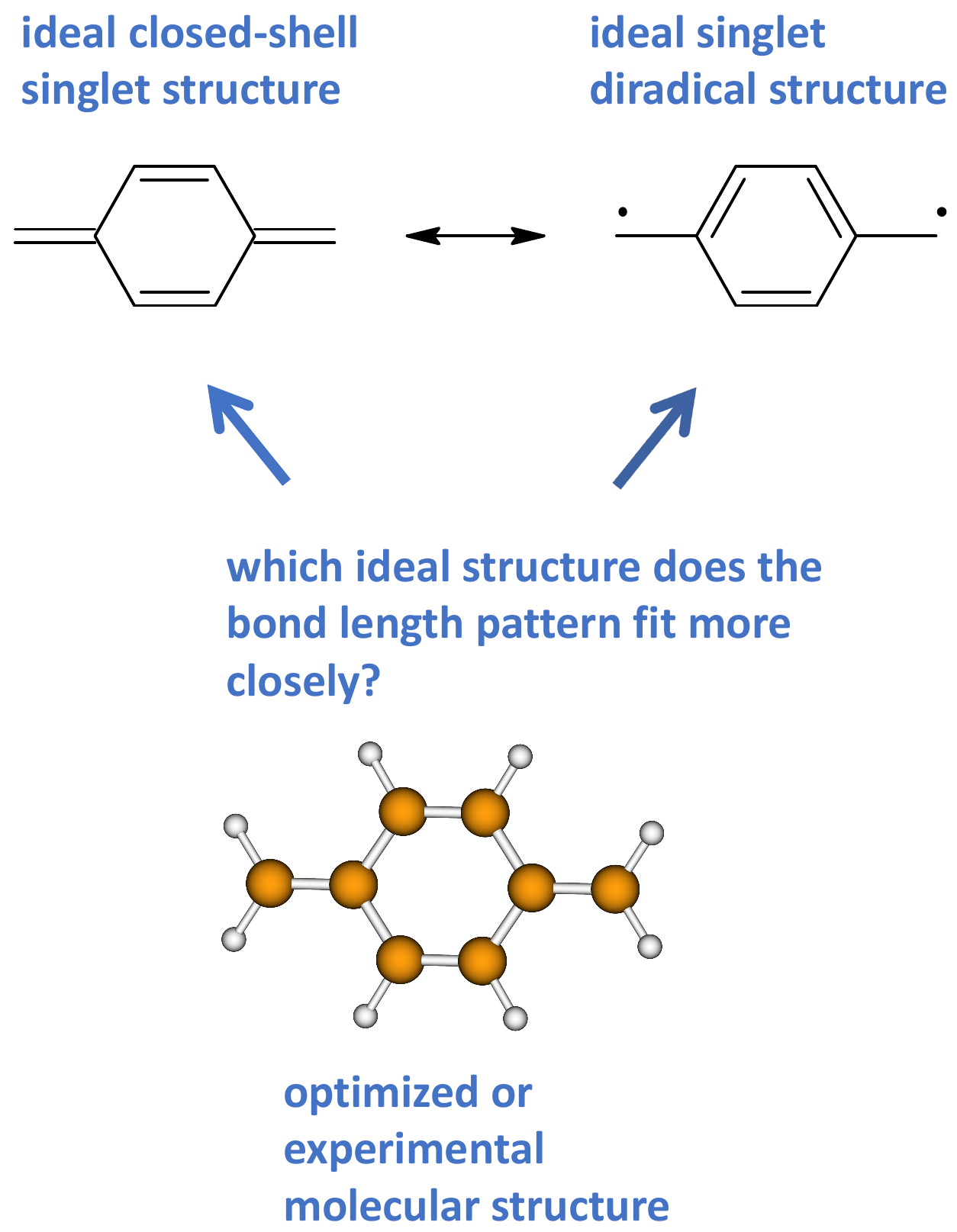}
	\caption{\textcolor{comm}{Our measure of structural diradical character is based on comparing bond-length patterns of molecular structures with idealized bond-length patterns for closed- (top left) and open-shell diradical (top right) forms, shown here for p-quinodimethane. 
}}
	\label{fig:pqm}
\end{figure}

\textcolor{comm3}{The singlet state of diradicals can have a bond length pattern 
more reminiscent of an open-shell structure (Figure~\ref{fig:pqm}, top right),
of a closed-shell structure (Figure~\ref{fig:pqm}, top left), or somewhere 
in between. A perfect open-shell
molecular structure will typically be very close to that of a triplet. 
Depending on which side the molecular structure leans to, electronic
properties will be considerably different, in particular the (electronic) diradical
character.} 
Indeed, it has been shown that  
diradical character and NLO properties can be very sensitive to molecular structure~\cite{blaex,vila10,seid14,sher92}. Open-shell electronic structures have also \textcolor{comm}{been} found to depend on interatomic 
distances in the context of strongly correlated adsorbates and materials~\cite{bahl18,Skornyakov2017,Plihal1998}.
\textcolor{comm3}{For predicting diradical properties from first principles, 
it is therefore important to predict sufficiently accurate molecular structures, both in terms of absolute bond lengths and in terms of bond length patterns.} 


\begin{figure}[ht]
	\centering
		\includegraphics[width=0.89\textwidth]{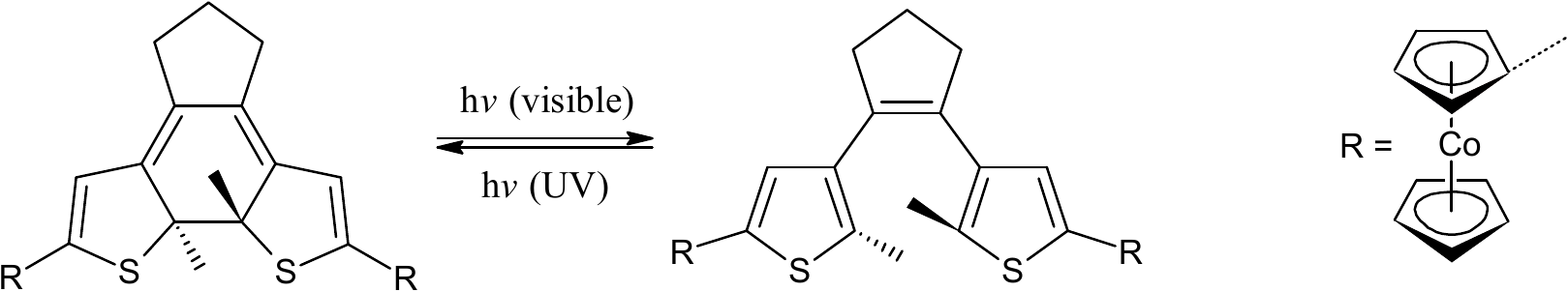}
	\caption{Lewis structures of a dithienylethene molecule in its closed (left) and open form (right).}
	\label{fig:dte}
\end{figure}


\textcolor{comm3}{For many diradicals of interest, \ac{KS}-\ac{DFT} is the only 
first-principles electronic structure method capable in practice 
of molecular structure 
predictions with reasonable accuracy (see also Appendix~\ref{sec:bs}).
Yet, owing to the unknown exchange--correlation functional, \ac{DFT} can give
inconclusive results regarding such structures.}

\begin{figure}[ht]
	\centering
		\includegraphics[width=1.00\textwidth]{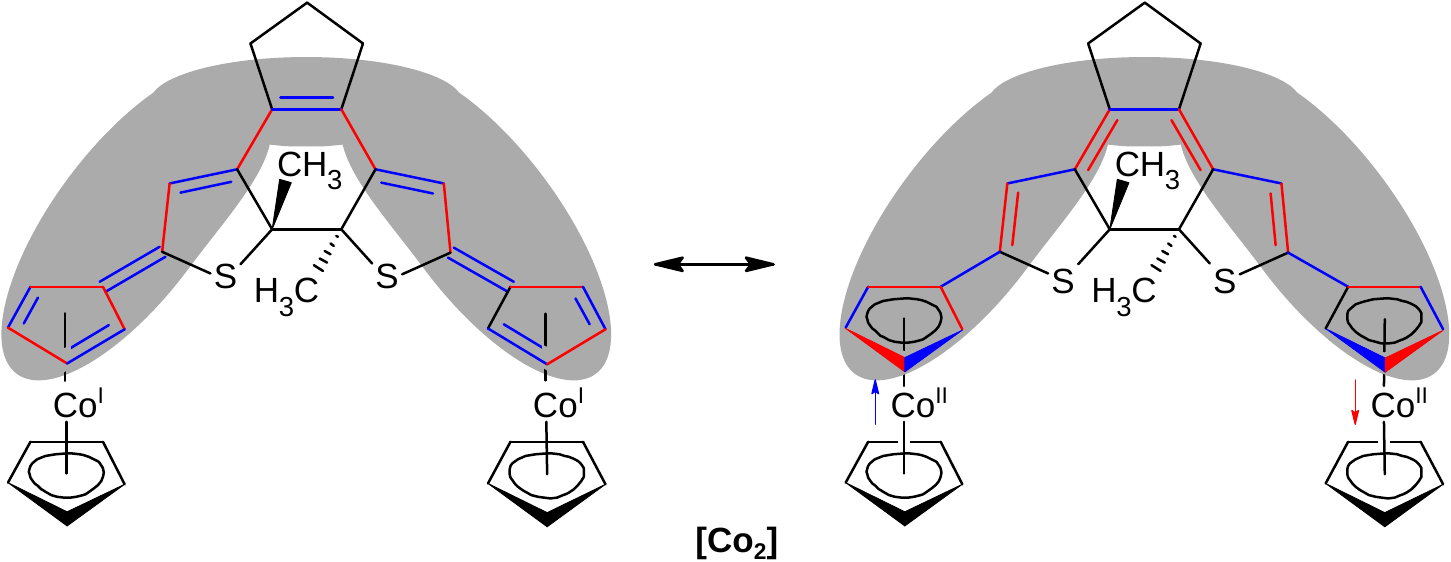}
	\caption{Lewis structures of \textbf{[Co\textsubscript{2}]}. The closed-shell quinoidal form is shown on the left-hand side, the diradicaloid form on the right-hand side. Note that strictly speaking, these are not mesomeric forms, since bond lengths will differ between the diradicaloid and the closed-shell structures. The bonds whose formal bond orders and thus lengths differ between the two structures  are indicated by the grey area.
Bonds included in evaluating established \ac{BLA} measures are shown in red if they were added and in blue if they were subtracted in Eq.\ (\ref{eq:bla})~\cite{bladef1,bladef2}.}
	\label{fig:switch}
\end{figure}

\textcolor{comm3}{ We are interested in 
a particular example of such inconclusive predictions, a dithienylethene-linked 
biscobaltocene whose diradical properties could be switched, in principle,
by light (see Figure~\ref{fig:dte}). 
In combination with the redox-active nature of the cobaltocene units,
this might lead to multiresponse behavior}.
Closing the photoswitch (left-hand side
of Figure~\ref{fig:dte}) switches on electronic communication via the
$\pi$ system of the bridge, which enables drawing two \textcolor{comm}{different} structures, 
a closed-shell \textcolor{comm3}{one (Figure~\ref{fig:switch}, left) and an open-shell one (Figure~\ref{fig:switch}, right)}. 
\textcolor{comm2}{The relative importance of these two not only affects 
 \ac{NLO} properties, but a stabilization of the closed switch resulting
from a large admixture of the closed-shell form can also 
suppress photochromic ring opening~\cite{mats00b,kawa94}.}
Due to its poor switching behavior, the closed-switch form could not be isolated experimentally,
and its structure and properties are therefore not known yet~\cite{escribano2017}.
To decide whether further efforts towards obtaining these data and 
towards optimizing the switching behavior are worthwhile,
we aim at a true first-principles prediction of the diradical character
of the closed switch.
In contrast to the analogous nitronyl nitroxide compound~\cite{mats00} (see Supplemental Material), 
\ac{KS}-\ac{DFT} optimizations of the molecular structure for the closed switch 
in its singlet state give no consistent answer to whether it is predominantly a 
closed-shell or an open-shell structure \textcolor{comm3}{(see Section~\ref{sec:co2})}. 
Consequently, no predictions of its 
diradical character and its \ac{NLO} properties appear possible, unless a
particular approximate exchange--correlation functional can be identified 
as sufficiently reliable for this purpose.

\begin{figure}[ht]
	\centering
		\includegraphics[width=1.00\textwidth]{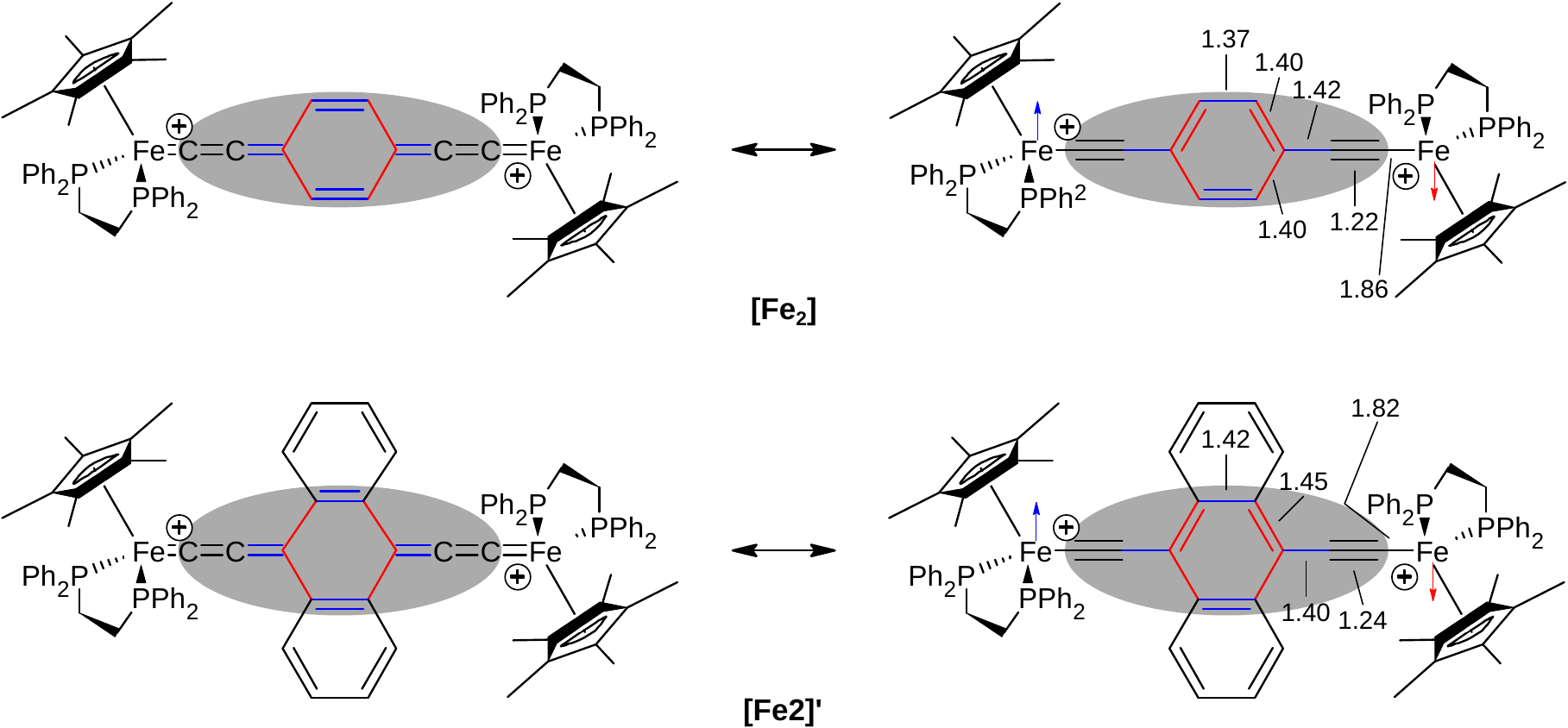}
	\caption{Lewis structures of \textbf{[Fe\textsubscript{2}]} and \textbf{[Fe\textsubscript{2}]'}. The bond lengths (in {\AA}) for \textbf{[Fe\textsubscript{2}]} and \textbf{[Fe\textsubscript{2}]'} are taken from X-ray crystal structures from Ref.~\cite{lapinte_orig} and~\cite{lapinte}, respectively. Note that the shown structures are, strictly speaking, not resonance structures, since they have different bond lengths. Bonds included in evaluating established \ac{BLA} measures are shown in red if they were added and in blue if they were subtracted in Eq.\ (\ref{eq:bla})~\cite{bladef1,bladef2}.
Note that the reference bond lengths within the benzene rings were those of 
aromatic benzene and not the alternating single and double bonds shown in the 
Lewis structures. Also, the \ce{Fe-C} bonds are not included in the evaluation of the structural diradical character, because no reference bond length for \ce{Fe-C} single and \ce{Fe=C} double bonds were defined.
	}
	\label{fig:lapinte-structures}
\end{figure}

\begin{figure}[ht]
	\centering
		\includegraphics[width=1.00\textwidth]{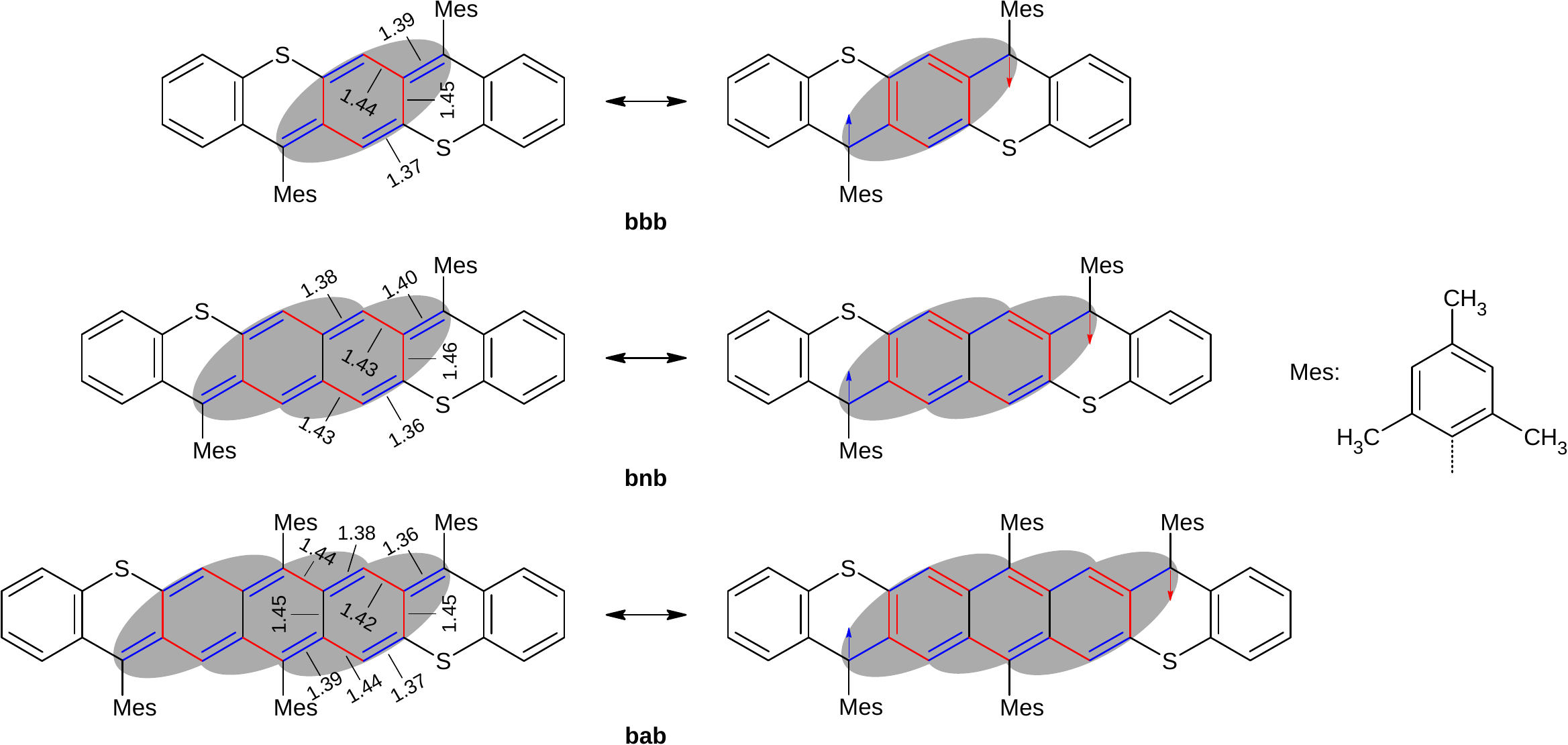}
	\caption{Lewis structures and selected X-ray crystallographic bond lengths (from Ref.~\cite{bisbenzothia}) of \textbf{bbb}, \textbf{bnb} and \textbf{bab}. The closed-shell quinoidal form is shown on the left-hand side, the diradicaloid form on the right-hand side. Note that strictly speaking, these are not mesomeric forms, since bond lengths will differ between the diradicaloid and the closed-shell structures. Bonds included in evaluating established \ac{BLA} measures are shown in red if they were added and in blue if they were subtracted in Eq.\ (\ref{eq:bla})~\cite{bladef1,bladef2}.}
	\label{fig:bbtqdm-structures}
\end{figure}

\textcolor{comm3}{One would expect that \acf{BLA}~\cite{bladef1,bladef2} 
was a good measure for comparing molecular structures optimized with
different approximate exchange--correlation functionals and experimental structures.
It turns out that \ac{BLA} values spread so unsystematically that 
this is not possible (along with other disadvantages, as discussed in Appendix~\ref{sec:bla}). Therefore, we will define a new measure 
for this purpose, which we call structural diradical character (see Section~\ref{sec:dirad}). It is based
on measuring the deviation of bond lengths for a structure of interest from
ideal bond lengths of (a) an open-shell singlet and (b) a closed-shell singlet (see Figure~\ref{fig:pqm}), 
obtained with the same methodology as the structure of 
interest (DFT with a particular exchange--correlation functional, or experiment).
We  will show below that this indeed allows for identifying functionals 
that can be considered reliable for bond length patterns of singlet diradicals.
For this purpose we will analyze a series of experimentally studied diradicals (see Figures~\ref{fig:lapinte-structures} and~\ref{fig:bbtqdm-structures}).
 }


\section{Structural diradical character}
\label{sec:strucdirad}

Diradical character is a measure used to indicate how close a system resembles one with two unpaired electrons (usually in a singlet state).
Open-shell character is a more general term that can be used in various contexts.
It may refer to diradical character (two unpaired electrons) as well as to any other polyradical character (any number of unpaired electrons).
Here, we use the terms diradical character and open-shell character synonymously.
\subsection{Defining structural diradical character: How close is the bond length pattern of a molecular structure to that an ideal diradical?}
\label{sec:dirad}

We introduce a new measure for estimating the qualitative similarity of a molecular structure to an ideal diradical or closed-shell bond pattern (see Figure~\ref{fig:pqm}). 
The new measure overcomes the drawbacks of the \ac{BLA} 
scheme while still retaining its simplicity. 
For this purpose, reference bond lengths for the ideal open-shell and closed-shell structures have to be defined (see Section~\ref{sec:reference} below for details). 
The actual bond lengths $b$ for the structure of interest 
are then 
compared to these reference bond lengths $b_{\rm ref}$, and the normalized mean absolute error ($\mathrm{MAE}_{\mathrm{norm}}^{X}$; $X=\mathrm{CS},\mathrm{OS}$),

\begin{equation}
\mathrm{MAE}_{\mathrm{norm}}^{X} = \frac{\sum_{i=1}^{n} \frac{|b_{i}-b_{i,\mathrm{\rm ref}}^{X}|}{|b_{i,\mathrm{\rm ref}}^{\mathrm{OS}}-b_{i,\mathrm{\rm ref}}^{\mathrm{CS}}|}}{n},
\label{eq:nmae}
\end{equation}

and the normalized root mean squared deviation ($\mathrm{RMSD}_{\mathrm{norm}}^{X}$; $X=\mathrm{CS},\mathrm{OS}$),

\begin{equation}
\mathrm{RMSD}_{\mathrm{norm}}^{X} = \sqrt{\frac{\sum_{i=1}^{n} \left( \frac{b_{i}-b_{i,\mathrm{ref}}^{X}}{|b_{i,\mathrm{ref}}^{\mathrm{OS}}-b_{i,\mathrm{ref}}^{\mathrm{CS}}|} \right)^{2}}{n}},\label{eq:nrmsd}\end{equation}

are calculated for all $n$ bonds where the reference bond length in the \ac{CS} and \ac{OS} differ (these bonds are encapsulated in gray ovals in the corresponding Figures). The normalization is used to account for different magnitudes in reference bond-length differences (\SI{20}{\pico\meter} for a \ce{C-C -> C=C} transition, \SI{14}{\pico\meter} for a \ce{C=C -> C#C} transition, \SI{14}{\pico\meter} for a \ce{C-C -> C\bond{~-}C} transition and \SI{6}{\pico\meter} for a \ce{C=C -> C\bond{~-}C} transition). 

The structural diradical character $y_{s}$ is then defined as

\begin{equation}
y_{s} = 1 - \frac{\mathrm{MAE}_{\mathrm{norm}}^{\mathrm{OS}}}{\mathrm{MAE}_{\mathrm{norm}}^{\mathrm{OS}}+\mathrm{MAE}_{\mathrm{norm}}^{\mathrm{CS}}} = \frac{\mathrm{MAE}_{\mathrm{norm}}^{\mathrm{CS}}}{\mathrm{MAE}_{\mathrm{norm}}^{\mathrm{OS}}+\mathrm{MAE}_{\mathrm{norm}}^{\mathrm{CS}}}.
\label{eq:osc}
\end{equation}

The structural diradical character 
can be calculated from the $\mathrm{RMSD}_{\mathrm{norm}}^{X}$ in an analogous way.

\subsection{Choosing reference bond lengths for diradical and closed-shell structures}
\label{sec:reference}

It is not quite obvious how ideal diradical and closed-shell singlet bond lengths should be defined. 
One option would be to carry out a computational structure optimization in which the electronic structure is constrained to be a closed-shell singlet (as in spin-restricted KS-DFT) and to use the resulting bond lengths as references for the closed-shell structure, and, accordingly, a spin-unrestricted triplet optimization for a ``perfect'' open-shell structure. This has the obvious disadvantage that it cannot be applied to experimental structures.

Another option is to interpret the Lewis structures literally and to use \ce{C-C} bond lengths of ethane, ethene, ethine, and benzene, \dots as references. 
These ideal bond lengths are obtained with the same method as the molecular structure of interest, i.e., either by structure optimization with a method like DFT  (see Supplemental Material for values), or from tabulated experimental data for these validation compounds (taken from Ref~\cite{bond-lengths}) \cite{bla1}. 
The drawback of this definition is that a ``real-world'' open-shell
singlet may by be quite far from
a  perfect bond pattern as shown on the left-hand side of Figure~\ref{fig:pqm},
because \textcolor{comm2}{$\pi$ conjugation~\cite{crai06}},
chemical substitution, intramolecular and intermolecular dispersion interactions and repulsions, and other effects may lead to deviations from ideal
bond lengths. For the same reasons, what we would clearly consider a closed-shell
singlet structure may deviate from its ideal bond pattern as shown on the right-hand side of Figure~\ref{fig:pqm}. 
We do not consider this
a major problem, because these reasons are present in computationally optimized and experimental structures alike.  We do acknowledge that (1) in practice, intermolecular interactions are usually neglected in DFT (but this is the case for nearly all computational work) and (2) in cases where a certain exchange--correlation functional has a weakness concerning, e.g., intramolecular dispersion interactions indirectly affecting bond length patterns, we may not be able to disentangle intrinsic problems of this functional with bond length patterns from its weaknesses with dispersion. However, with these exceptions, we consider comparing the structural diradical character between experiment and computation as a valuable means of gaining insight into the reliability of electronic structure methods for molecular structure optimizations of diradicals. 


\FloatBarrier

\section{An illustrative example: Electronic diradical character depends on bond length patterns much more than on absolute bond lengths}
\label{sec:example}

The studied molecules can be drawn in different forms (see Figures~\ref{fig:switch},~\ref{fig:lapinte-structures} and~\ref{fig:bbtqdm-structures}), one of which denotes a \ac{CS} and the other an \ac{OS} form bearing two unpaired electrons.
These two forms differ in the distribution of \ce{C-C} bonds (single-, double-, triple- and aromatic bonds) between respective carbon atoms.
In the \ac{CS}, the bond lengths are typically more equally distributed than in the \ac{OS}.
These bond length distributions or bond length patterns can be used to evaluate the structural diradical character.

\begin{figure}[ht]
        \centering
                \includegraphics[width=1.05\textwidth]{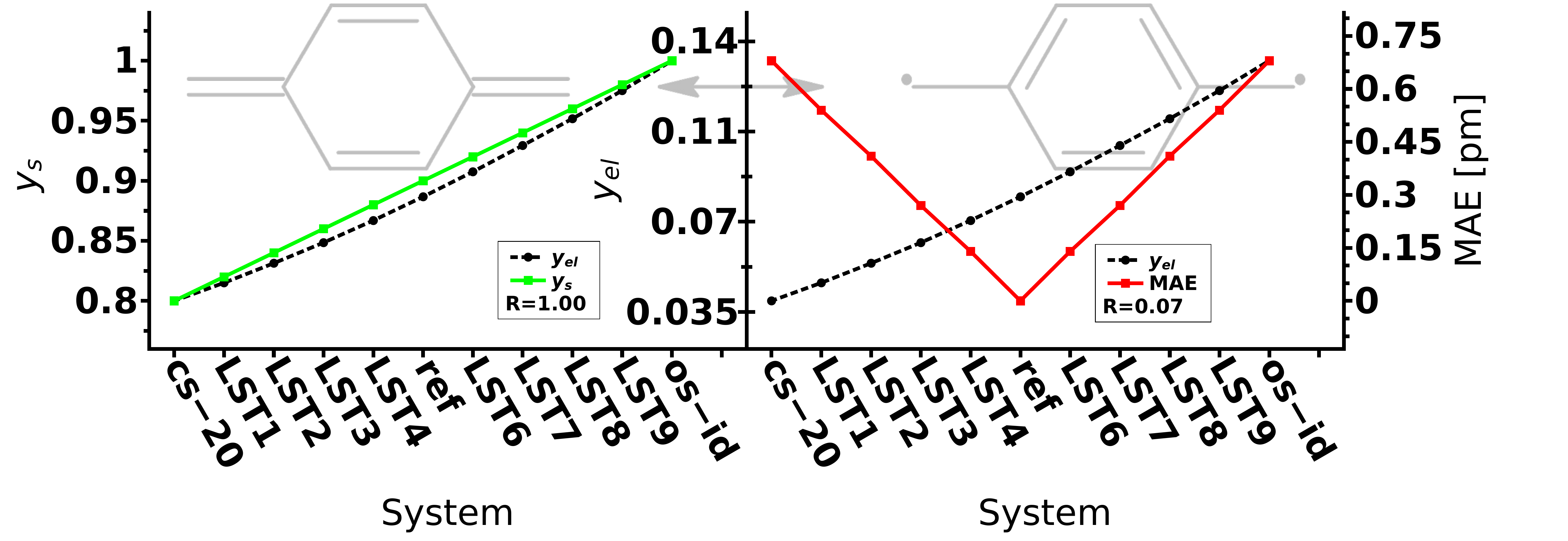}
        \caption{Comparison of the correlation between electronic ($y_{el}$) and structural ($y_{s}$) diradical character ($\mathrm{MAE}$, Equation~\eqref{eq:osc}) on the left (right). The Pearson correlation coefficient $R$ is shown in the corresponding legends. The systems are p-quinodimethanes with different bond lengths. cs-20 ($\text{ref}$, os-id) is a linear combination of $20$~\% ($10$~\%, $0$~\%) cs-id and $80$~\% ($90$~\%, $100$~\%) os-id, the structures denoted as $\text{LST}i$ are linear combinations of $20-2i$~\% cs-id and $80+2i$~\% os-id. To make a qualitative comparison easier, the scales of the ordinates were chosen so as to have the smallest (and largest) data points of the respective curves on equal heights. Single point calculations used B3LYP/def2-TZVP-D3.}
        \label{fig:pqmcomp}
\end{figure}

{We will show in the following 
that the structural diradical character correlates more with electronic diradical character\cite{blaex,naka17,ramo14,bach02,here03,brai17,nees04} than the $\mathrm{MAE}$ of absolute bond lengths does.
To illustrate this (see Figure~\ref{fig:pqmcomp}), we have compared the correlation of structural diradical character and $\mathrm{MAE}$ of a set of different structures of p-quinodimethane.
}

{
First, we describe the construction of the studied set of structures.
We take two structures as end points for a linear interpolation:
First, a structure resembling the closed-shell Lewis structure was built and its bond lengths were chosen to be the ideal bond lengths that are used as a reference in the calculation of the structural diradical character (see Table~S3 in the Supporting Information).
This structure then has a structural diradical character of $0$.
We will call this structure ideal closed shell or cs-id.
The second structure was constructed in a similar fashion, but this time, the ideal bond lenghts resembling the open-shell Lewis structure were chosen, resulting in a structural diradical character of $1$.
Analogously, this structure is referred to as ideal open shell or os-id.
The $11$ studied structures (see Section~S8 in the Supporting Information for cartesian coordinates) were then built as linear interpolations between cs-id (with weights between $0.0$ and $0.2$, the latter being referenced as cs-20 $=0.2 \cdot $cs-id + $0.8 \cdot$os-id) and os-id (with weights between $1.0$ and $0.8$).
The weights were chosen to be in a region where the electronic diradical character is sensitive to structural changes.
We computed the electronic diradical character $y_{el}$, structural diradical character $y_{s}$,
and the $\mathrm{MAE}$ of the bond lengths that were used for calculating $y_{s}$ (the \ce{C-C} bond lengths) with respect to a fititious ``validation'' 
structure which is a linear combination of os-id (weight is $0.9$) and cs-id (weight is $0.1$), denoted as $\text{ref}$.
The latter structure corresponds to what would usually be a molecular
structure from the experiment, and was
chosen such that it is in the middle of the range of structural mixtures under study.}

\begin{figure}[ht]
        \centering
                \includegraphics[width=0.75\textwidth]{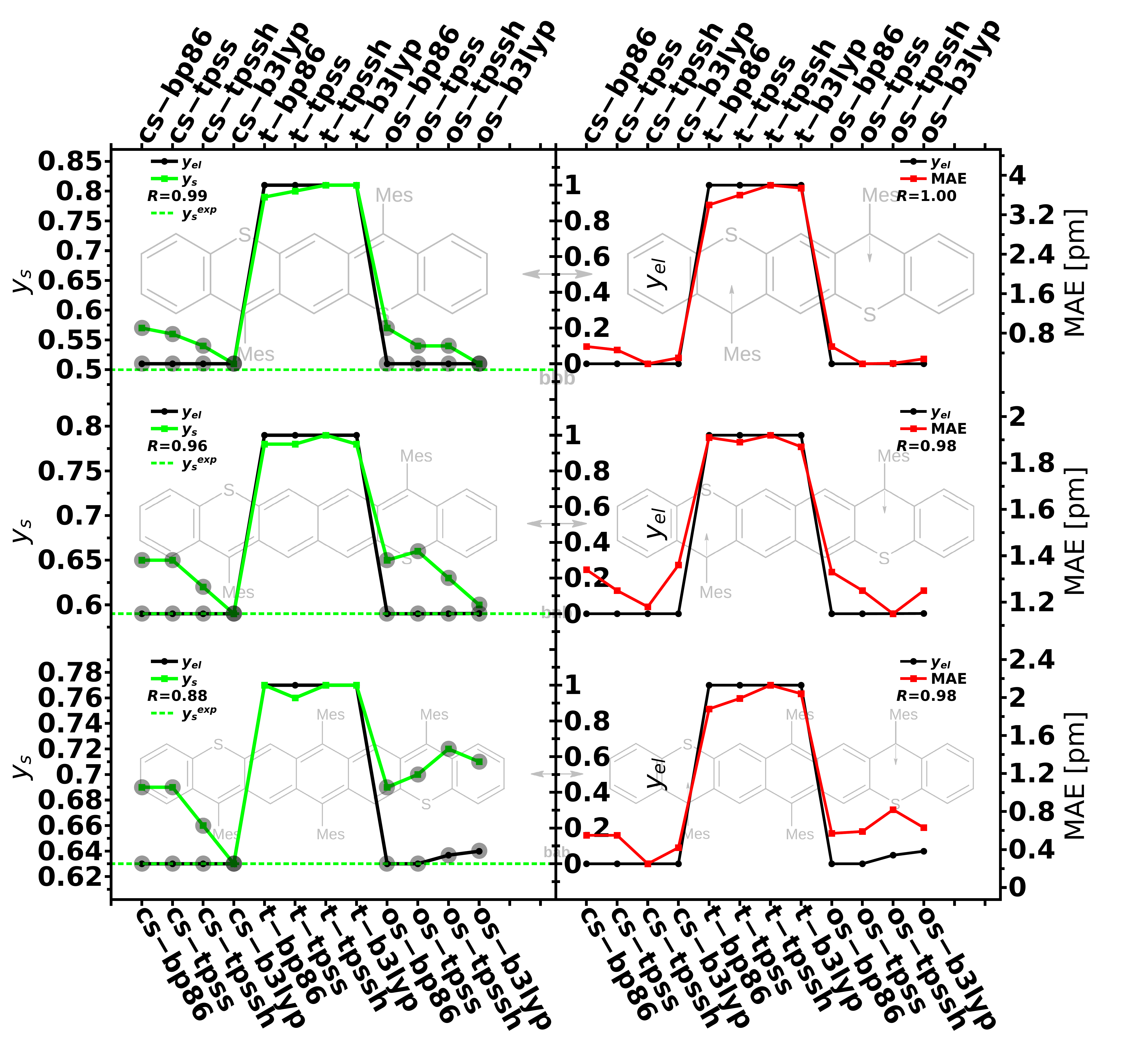}
        \caption{Comparison of the correlation between electronic diradical character $y_{el}$ and structural diradical character $y_{s}$ ($\mathrm{MAE}$) on the left (right).
MAE values are calculated from geometry optimized- and experimental structures.
The Pearson correlation coefficient $R$ is shown in the corresponding legends. The data for \textbf{bbb} are shown in the top third, for \textbf{bnb} are shown in the center third and for \textbf{bab} are shown in the bottom third.
Values for $y_{el}$, $y_{s}$ and $\mathrm{MAE}$ are plotted against a string representing the calculated determinant (either cs for closed shell, t for triplet or os for broken symmetry) and the used \ac{xc} functional.
Additionally, the calculated structural diradical character values for the x-ray structures $y_{s}^{exp}$ are plotted as a constant dotted line. The energetically most stable structures are highlighted by a circular grid on top of the data point. Again, energies differing by less than $5$~kJ are considered degenerate, leading to multiple highlighted structures per functional in some cases.}
        \label{fig:bbtqdm-comp-ys-yel-mae}
\end{figure}

{The better correlation between electronic and structural diradical character than between electronic diradical character and the $\mathrm{MAE}$ is evident.
The Pearson correlation coefficient between $y_{s}$ and $y_{el}$ (Figure~\ref{fig:pqmcomp}, left) is $1.00$, while it is $0.07$ between $\mathrm{MAE}$ and $y_{el}$ (Figure~\ref{fig:pqmcomp}, right).
For the realistic systems under study, we will show that structural diradical character shows a similarly better correlation with electronic diradical character than $\mathrm{MAE}$ (see Figures~\ref{fig:lapinte-comp-ys-yel-mae} and~\ref{fig:bbtqdm-comp-ys-yel-mae}).}

Accordingly, correct bond length patterns are more important for getting electronic diradical character right than only good agreement with absolute bond lengths. In particular, good agreement with absolute bond lengths may lead to considerable deviations in electronic diradical character if it was obtained at the expense of realistic bond length patterns.
We therefore suggest a measure for the agreement with bond length patterns as an important additional criterion when evaluating the performance of electronic structure methods for molecular structure optimizations of diradicals, in addition to measures for absolute bond length deviations.


\section{Attempt at a true first-principles prediction of diradical character: biscobaltocenyldithienylethene}
\label{sec:co2}

\ac{DTE} derivatives can, in principle, be switched from a closed form (Figure~\ref{fig:dte} left), which has an extended conjugation, to an open form (Figure~\ref{fig:dte} right)
by radiation with visible light. Ring closure can be initiated by irradiation with UV light (due to a less extended conjugation in the open form). For the closed form with the attached cobaltocenes (\textbf{[Co\textsubscript{2}]}), a diradical Lewis structure with two unpaired electrons and a closed-shell structure with no unpaired electrons can be drawn (see Figure~\ref{fig:switch}). It is not known yet experimentally 
whether the molecule is predominantly \ac{OS} or \ac{CS} in its ground state. 

In an attempt to make a true first-principles prediction on 
this question, we have calculated optimized structures 
(open and closed shell) and analyzed the spin-state energetics 
and structural diradical characters of this compound, employing 
the pure BP86 functional and the hybrid B3LYP with 20 \% admixture of
Hartree--Fock exchange. 
We start molecular structure optimizations for (1) a closed-shell singlet (cs), employing 
spin-restricted \ac{KS} \ac{DFT} (RKS), (2) a broken-symmetry~\cite{nood81} (bs) approximation
of the open-shell singlet  employing 
spin-unrestricted \ac{KS} \ac{DFT} (UKS), for which the local spin density 
in the initial guess corresponded to one spin-up unpaired electron on
one spin center and one spin-down electron on the other 
center~\cite{n3}, and (3) a triplet (t) described by UKS to evaluate 
singlet--triplet splittings. 
For a bs solution with approximately one unpaired electron per spin center,
a total spin expectation value of $\langle \hat S^2\rangle$ close to one 
is expected~\cite{szab96}.
The bs approach may converge to a closed-shell solution, which is indicated
by $\langle \hat S^2\rangle$ approaching zero. Therefore, $\langle \hat S^2\rangle$ 
values are reported for all cs and bs optimizations. 
All molecular structures under study have singlet ground states.

\begin{table}[ht]
\centering
\caption{Relative energies with respect to the closed-shell energy ($\Delta E$) [kJ/mol],
structural \textcolor{comm2}{diradical} character $y_{s}$ (Equation~\ref{eq:osc})  (MAE and, in parentheses, RMSD),
bond-length alternation (Equation~\ref{eq:bla}) $\mathrm{BLA}$ [pm],
$\hat{S}^{2}$ expectation values ($\braket{\hat{S}^{2}}$)
for  the optimized structures of
\textbf{[Co\textsubscript{2}]} 
(see Figure~\ref{fig:switch}) employing BP86/def2-TZVP and B3LYP/def2-TZVP,
for closed-shell (cs), open-shell singlet modeled by a broken-symmetry (bs) determinant,
and triplet (t). The structural diradical characters of the energetically most stable structures are highlighted in green (with energies differing by less than 5 kJ/mol considered as degenerate). The overall assignment as closed-shell (CS) or open-shell (OS) is indicated in the right-most column.}
\label{tab:switch-complete}
\begin{tabular}{|c||c|c|c|c||c|c|c|c||c|c|c||c|}
\hline
\cellcolor[HTML]{C0C0C0} & \multicolumn{4}{c||}{cs} & \multicolumn{4}{c||}{bs} & \multicolumn{3}{c||}{t} &\\ 
\cline{2-13}
\cellcolor[HTML]{C0C0C0} & $\Delta E$ & $y_{s}$ & $\rm{BLA}$ & $\braket{\hat{S}^{2}}$ & $\Delta E$ & $y_{s}$ & $\rm{BLA}$ & $\braket{\hat{S}^{2}}$ & $\Delta E$ & $y_{s}$ & $\rm{BLA}$ &  \\ \hline
\cellcolor[HTML]{C0C0C0} & \multicolumn{11}{c}{\textbf{BP86}} & \\ 
\cline{2-13}
\textbf{[Co\textsubscript{2}]} & 0.00 & \textcolor{Green}{0.58 (0.71)} & 2.6 & 0.00 & 1.37 & \textcolor{Green}{0.58 (0.71)} & 2.6 & 0.01 & 28.45 & 0.67 (0.81) & -1.43 & {CS} \\ \hline
\cellcolor[HTML]{C0C0C0} & \multicolumn{11}{c}{\textbf{B3LYP}} & \\ \cline{2-13}
\textbf{[Co\textsubscript{2}]} & 0.00 & 0.56 (0.67) & 4.4 & 0.00 & -52.6 & \textcolor{Green}{0.73 (0.87)} & -3.4 & 1.10 & 2.04 & 0.74 (0.88) & -3.7 & {OS} \\ \hline
\end{tabular}
\end{table}
For all optimized structures, we 
evaluate structural diradical character and \ac{BLA}.
If these data agreed reasonably well for the ground-state structures of 
both BP86 and B3LYP,
a DFT prediction of the open-shell character of \textbf{[Co\textsubscript{2}]} 
could be considered as reliable. 
However, BP86 gives a closed-shell structure, 
while B3LYP results in an open-shell singlet as the energetically 
most stable solution (Table~\ref{tab:switch-complete}).

The structural diradical character is small for the closed-shell solutions and large for the open-shell solution, which could only be converged with B3LYP, not with BP86, even if performing a single-point calculation on the open-shell optimized B3LYP structure with the open-shell B3LYP orbitals as initial guess.  Interestingly, the BP86 closed-shell singlet shows a larger structural diradical character than the B3LYP one, indicating that even though both are equally closed-shell in terms of electronic structure ($\braket{\hat{S}^{2}}=0$), the BP86 
molecular structure leans more towards the open-shell side. \textcolor{comm2}{For comparison, structural diradical character was also evaluated for the triplet. Here, one would expect
bond length patterns close to the open-shell resonance structures, and accordingly, 
$y_s$ is always largest for the triplet. For the B3LYP open-shell singlet, $y_s$
is nearly identical to the value for the triplet, which could be taken as an additional
indication that the open-shell-singlet B3LYP solution converges to a nearly pure diradical.}

Here, MAE- and RMSD-derived structural diradical characters 
deviate by up to 0.14, with the latter being larger, while in our validation systems (see below), 
it does not matter much whether 
structural diradical character $y_s$ is evaluated with MAE or RMSD as a measure 
for structural deviations, and the latter is typically slightly smaller.
 This different behavior might be related to the
fact that single / double bond alternation plays a more pronounced 
role here in \textit{both} structures in contrast to the validation compounds.
This does not affect the suitability of $y_s$ for comparisons between 
calculated and experimental data. It
makes employing $y_s$ as an absolute measure for diradical 
character difficult, but as will be discussed below, 0.6 as evaluated
based on MAE appears to be a reasonable measure for the transition
from what is typically considered more closed-shell to 
more open-shell singlet, at least for 
the set of molecules considered here.

A negative \ac{BLA} would be expected for an open-shell structure, because the single bonds of the bridge are subtracted (blue in Figure~\ref{fig:switch}) and the double bonds (shorter) are added (red in Figure~\ref{fig:switch}) and the aromatic bonds (same bond lengths in the reference) will not bias the \ac{BLA} towards positive or negative values. On the other hand, a positive \ac{BLA} indicates a closed-shell structure, because then the single bonds are added (red in Figure~\ref{fig:switch}), while the double bonds (shorter) are subtracted (blue in Figure~\ref{fig:switch}). The \ac{BLA} (Equation~\ref{eq:bla}) obtained from the BP86 solutions is small, but positive, rather corresponding to a closed shell, while the \ac{BLA} obtained from B3LYP is positive for the closed-shell solution and negative for the open-shell solution.

In accordance with the larger open-shell character suggested \textcolor{comm}{by} B3LYP, 
the singlet--triplet gap is by more than an order of magnitude smaller
than the gap predicted by BP86.

Altogether, these data suggest that according to BP86, \textbf{[Co\textsubscript{2}]} is 
mostly a closed-shell molecule, while B3LYP suggests it is mostly open-shell.
Therefore, in the following we will compare these two \ac{xc} functionals employed along
with two meta-GGA based ones (TPSS and TPSSh) with experimental data on structures where 
varying the bridge modifies the diradical character.

\FloatBarrier

%
%

\section{Comparison of DFT with experimental molecular structures}
\subsection{
Selection of diradicals and exchange--correlation functionals}
\label{sec:study}

For organic diradicals, the B3LYP exchange--correlation functional \textcolor{comm2}{(with 20 \% exact exchange admixture)} generally
works well~\cite{shil18}, but even here,  \textcolor{comm2}{\ac{BLA}~\cite{bladef1,bladef2}} as present in the closed-shell form on 
the left-hand side of Figure~\ref{fig:pqm} can be underestimated~\cite{neno13} or overestimated~\cite{seid13,seid14}, depending on the molecule studied.
While numerous studies on the dependence of NLO properties on exchange--correlation functional have been carried out~\cite{vila10,cham00, bula05,seki07,jacq07,supo08,bori09,kara11,sun13, garr14,garz15,wang17,kish10,buec14,ye07,ye06}, it is not 
clear if there is a reasonably reliable functional for 
describing  \textcolor{comm2}{bond length patterns} 
for organometallic complexes with potential diradical character.

We therefore apply the structural diradical measure defined in Section~\ref{sec:strucdirad} 
to two sets of 
selected organometallic~\cite{lapinte,lapinte_orig} and organic~\cite{bisbenzothia} 
validation compounds for which 
structural  data are
available from the experiment for given spin centers
with two and three different bridges, respectively,
and where these variations of the bridge are known to change the
diradical character considerably (see Figures~\ref{fig:lapinte-structures} and~\ref{fig:bbtqdm-structures}).
Even though the organometallic systems are quite large molecules, we consider,
in contrast to previous work\textcolor{comm}{\cite{lapinte,lapinte_orig}}, the full atomistic details of all ligands.

We compare four different exchange--correlation functionals,
three of which (BP86, TPSS, TPSSh) have proven valuable for structures and energetics 
of transition metal complexes~\cite{furch06,wall07,weym14}, while B3LYP is very popular for 
open-shell organic molecules. BP86 and TPSS are pure 
functionals, and TPSSh and B3LYP are hybrid functionals with 10 and 20 percent of 
exact-exchange admixture, respectively. The pure parts of BP86 and B3LYP are of
generalized gradient corrected (GGA) type, and for
TPSS and TPSSh, of meta GGA type.
Since exact exchange admixture tends to localize spin density~\cite{herr05,mmo,cram09}, hybrid functionals should favor diradical structures (right-hand side of Figure~\ref{fig:pqm})
more strongly than pure ones.

\subsection{Inorganic validation systems: Dinuclear carbon complexes with carbon-rich bridges}

In the two dicationic complexes shown in Figure~\ref{fig:lapinte-structures},
two iron(III) centers with one unpaired electron each are linked
by carbon-\textcolor{comm}{rich} bridges, in one case with a benzene linker (\textbf{[Fe\textsubscript{2}]}))
and in one case with a benzene linker featuring two annelated rings 
(\textbf{[Fe\textsubscript{2}]'}).
This annelation should decrease the aromaticity of the central carbon structure,
and thus favor the cumulenic structure shown on the left-hand side.

Indeed, in the experiment~\cite{lapinte}, 
analysis of characteristic bond lengths (\ce{Fe-C},\ce{-C#C},\ce{#C-C}) of \textbf{[Fe\textsubscript{2}]} revealed longer \ce{Fe-C} and \ce{#C-C} and shorter \ce{-C#C} bonds as compared to the X-ray structure of \textbf{[Fe\textsubscript{2}]'}, 
indicating a larger structural \textcolor{comm2}{diradical} character. 
Despite slight differences, this holds for both
molecules present in the unit cell (indicated by ``xray1'' ``and xray2'' in the lower part of Table~S4 in the Supporting Information)\cite{n2}.
Superconducting quantum interference device (SQUID) magnetometry (as powder) 
revealed a singlet ground state for \textbf{[Fe\textsubscript{2}]}
with a thermally accessible triplet state about 4.56 kJ/mol higher in energy,
suggesting that the ground state has significant open-shell character.
The \ac{VT}-UV--Vis spectra did not give any evidence of structural changes 
between 10~K and 300~K, and \ac{VT}-IR spectra pointed to a barely
detectable increase of cumulenic character with decreasing temperature,
suggesting that the structural changes between the singlet and triplet states
are minor. Altogether, this was interpreted as \textbf{[Fe\textsubscript{2}]}
having significant open-shell singlet character in its ground state,
despite the relatively strong antiferromagnetic spin coupling.

For \textbf{[Fe\textsubscript{2}]'}, no signal was found in the \ac{ESR} spectrum 
at 77~K, so the triplet state is not thermally accessible up to that temperature,
corresponding to a singlet--triplet splitting of at least 1200 cm$^{-1}$ (roughly 14.3 kJ/mol). This is also supported by \ac{VT}-NMR data. IR spectra support the
more cumulenic structure that was also indicated by X-Ray crystallography.
 This points to \textbf{[Fe\textsubscript{2}]'} having substantial closed-shell
character in the ground state.

\begin{figure}[ht]
        \centering
                \includegraphics[width=1.05\textwidth]{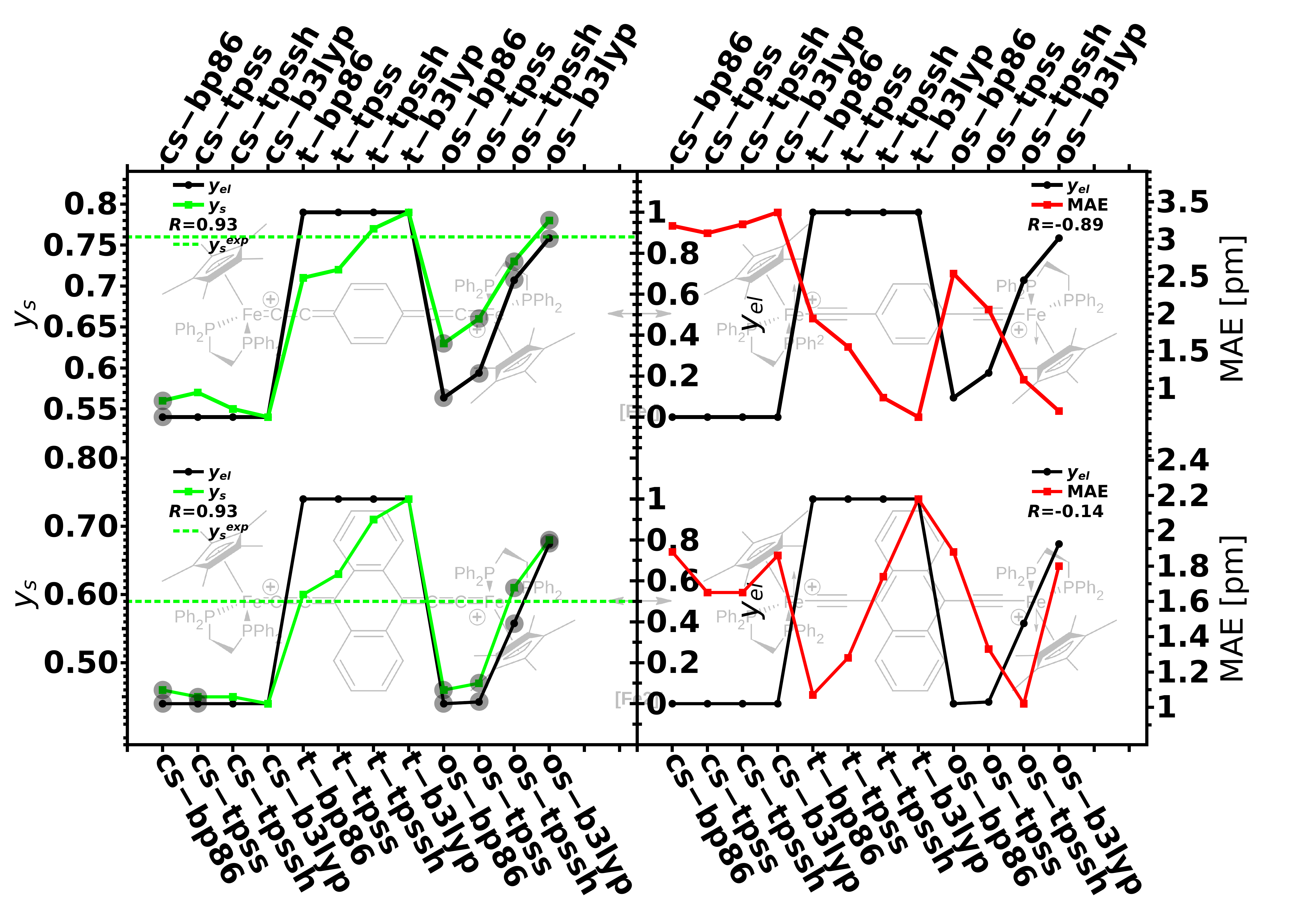}
        \caption{Comparison of the correlation between electronic diradical character $y_{el}$ and structural diradical character $y_{s}$ ($\mathrm{MAE}$) on the left (right).
$\mathrm{MAE}$ values are calculated from geometry optimized- and experimental structures.
The Pearson correlation coefficient $R$ is shown in the corresponding legends. The data for \textbf{[Fe\textsubscript{2}]} (\textbf{[Fe\textsubscript{2}]'}) are shown in the top (bottom) half.
Values for $y_{el}$, $y_{s}$ and $\mathrm{MAE}$ are plotted against a string representing the calculated determinant (either cs for closed shell, t for triplet or os for broken symmetry) and the used \ac{xc} functional.
Additionally, the calculated structural diradical character values for the x-ray structures $y_{s}^{exp}$ are plotted as a constant dotted line. The energetically most stable structures are highlighted by a circular grid on top of the data point. Again, energies differing by less than $5$~kJ are considered degenerate, leading to multiple highlighted structures per functional in some cases.}
        \label{fig:lapinte-comp-ys-yel-mae}
\end{figure}

{Figure~\ref{fig:lapinte-comp-ys-yel-mae} shows} that hybrid functionals,
in particular B3LYP and TPSSh, can describe the structural \textcolor{comm2}{diradical} character of 
\textbf{[Fe\textsubscript{2}]} very well {(compare the solid and dotted green lines)}. For \textbf{[Fe\textsubscript{2}]'},
B3LYP overestimates this character somewhat, while TPSSh is very close to
the experimental value (please see Section~S5 in the Supporting
Information for tabulated data). Interestingly, the reduction in $y_{s}$ from \textbf{[Fe\textsubscript{2}]} to \textbf{[Fe\textsubscript{2}]'} is partially described already by the
closed-shell-optimized structures. For the open-shell (bs) optimized structures,
$y_{s}$ increases as spins become more localized on the spin centers
(as indicated by increasing $\braket{\hat{S}^{2}}$ in Table~S4 in the Supporting Information). This is in line with this
localization indicating a stronger importance of the open-shell resonance structure (right-hand side of Figure~\ref{fig:lapinte-structures}).
\textcolor{comm2}{
The B3LYP open-shell-singlet structure for \textbf{[Fe\textsubscript{2}]} has a
 $y_{s}$ very close to the value for the triplet, which suggests that
B3LYP would consider \textbf{[Fe\textsubscript{2}]} almost purely
open shell. For TPSSh, the values are also reasonably close, indicating
dominant open-shell character.} 


Owing to the size of the systems under study
(we describe all ligands in full atomistic detail), 
crystal structure optimizations under periodic boundary conditions
(in particular with hybrid functionals) are prohibitively expensive.
It has been pointed out that 
crystal packing can increase quinoidal character~\cite{seid13}, so our 
first-principles 
structural diradical characters for isolated molecules may overestimate 
the values obtained from X-ray crystallography somewhat.
At least for \textbf{[Fe\textsubscript{2}]}, measured exchange spin coupling constants 
were very similar in the solid state and in solution~\cite{lapinte}, which suggests that 
structural differences between the two are not major. 
Still, it may be that the deviation of B3LYP data from the experiment 
is partially due to the neglect of packing effects.

Overall, based on 
the structural diradical character $y_s$, TPSSh would be considered adequate 
for describing the two iron-based complexes under study here (with B3LYP
being also acceptable). TPSSh also matches the 
experimental singlet--triplet energy splitting for \textbf{[Fe\textsubscript{2}]}
quite nicely (2.9 kJ/mol vs.\ 4.56 kJ/mol),
and is not too far from the experimental
lower bound on this splitting for \textbf{[Fe\textsubscript{2}]'}
(8.1 kJ/mol vs.\ 14.3 kJ/mol), whereas B3LYP underestimates both.

{In Section~\ref{sec:example}, we showed that the structural diradical character $y_{s}$ correlates with the electronic diradical character $y_{el}$ and that only taking into account averaged absolute bond length deviations $\mathrm{MAE}$ does not suffice
for a reliable comparison of computed and experimental data.
An analysis of the correlation between $y_{s}$ and $y_{el}$ and between $\mathrm{MAE}$ and $y_{el}$ has been conducted for the experimental validation systems as well (see Figure~\ref{fig:lapinte-comp-ys-yel-mae}).
Again, we see that structural diradical character correlates more strongly
with electronic diradical character than $\mathrm{MAE}$.
The correlation (expressed through the Pearson correlation coefficient $R$) is $0.93$ between $y_{s}$ and $y_{el}$  and $-0.89$ between $\mathrm{MAE}$ and $y_{el}$.
Here, strong anticorrelation between $\mathrm{MAE}$ and $y_{el}$ is observed, because the system under study is open shell.
There, one would expect the $\mathrm{MAE}$ between the experimental structure and the optimized one to be larger for the closed-shell structure (where $y_{el}$ is small) and smaller for open-shell structures (where $y_{el}$ is large).
The weak correlation between $\mathrm{MAE}$ and electronic diradical character at \textbf{[Fe\textsubscript{2}]'} is attributed to the higher level of complexity in the bonding patterns.
While in the organic systems, only alternations between single-, double- and aromatic bonds happen, while in the inorganic systems, alternations between single-, double-, aromatic- and triple bonds take place.
The structural diradical character can clearly deal with these complex bonding patterns.}

With bond-length alternation, it is more difficult to obtain a 
clear picture (see Supporting Information Table~S4).
Bond-length alternation should be more pronounced for the more quinoidal form
of \textbf{[Fe\textsubscript{2}]'} compared with \textbf{[Fe\textsubscript{2}]},
and this is indeed the case for the structures obtained from experiment. 
Also, the \ac{BLA} of an open-shell solution (if one is converged) is, 
as expected, smaller than for a closed-shell solution. 
For the DFT-optimized structures, \ac{BLA} data vary significantly 
depending on the \ac{xc} functional employed. 
The increase in \ac{BLA} from  \textbf{[Fe\textsubscript{2}]} to
 \textbf{[Fe\textsubscript{2}]'} is only reproduced for 
the two pure functionals (for which attempts at broken-symmetry optimizations
converge to $\braket{\hat{S}^{2}}$ smaller than one), and in terms of absolute numbers,
there is no functional which agrees well with the \ac{BLA} for both \textbf{[Fe\textsubscript{2}]} and \textbf{[Fe\textsubscript{2}]'}. For \textbf{[Fe\textsubscript{2}]}, B3LYP comes closest (but fails for \textbf{[Fe\textsubscript{2}]'}), and for 
\textbf{[Fe\textsubscript{2}]'}, TPSS matches best (but strongly 
overestimates \ac{BLA} for \textbf{[Fe\textsubscript{2}]}). 
Also, while $y_s$ changes only slightly when varying bond lengths within
the experimental error bar, these 
variations affect \ac{BLA} values considerably.
All this suggests that in contrast to $y_s$, 
it is at least difficult to identify a reliable
\ac{xc} functional for structural diradical character based on BLA.

\FloatBarrier

\subsection{Organic validation systems: bisbenzothiaquinodimethanes with varying bridge lengths}

Bisbenzothiaquinodimethanes (see Figure~\ref{fig:bbtqdm-structures}) 
have recently been presented as stable analogues of 
larger acenes, enabling the experimental study of diradical character 
as a function of molecular length~\cite{bisbenzothia}.
While all three molecules under study showed considerable quinoidal character,
diradical character increased with increasing molecular length as expected,
\textcolor{comm2}{owing to the increasing number of aromatic rings formed in
the open-shell resonance structure~\cite{malr16}}.
This was concluded, among others, from the X-Ray crystallographic structures and
from the increased spectral broadening in \ac{VT} \textsuperscript{1}H-NMR,
indicating more strongly thermally populated triplet states for longer molecules.

We optimized the molecular structures in the closed-shell, open-shell singlet (bs),
and triplet states with the four \ac{xc} functionals under consideration.
Here, we employed Grimme's empirical dispersion correction (D3)~\cite{grim11},
since it has been proven important for extended organic systems. 
We had not employed this correction for the inorganic complexes
above, because its suitability for inorganic systems is not as
clearly established as for organic ones~\cite{weym14}.

For all functionals and structures, we obtain either a closed-shell solution as the ground state, or an open-shell singlet (bs) that is close in energy to the closed-shell one. The energy differences between the two are at most around 3.5 kJ/mol, which appears too close to 
the DFT error margin to make a well-founded decision on which of the two represents
the ground state better. The  $\hat{S}^{2}$ expectation values of the bs solutions are 
often close to zero and never larger than about 0.6, indicating partial closed-shell character (see Table~S5 in the Supporting Information). The larger $\braket{\hat{S}^{2}}$, the more the structural diradical characters $y_s$ and 
bond-length alternation 
deviate from the ``true'' closed-shell solution. In all cases, the triplets 
are considerably higher in energy, consistent with the dominantly closed-shell ground states,
and the singlet--triplet splitting of 27 kJ/mol obtained from B3LYP-D3 is consistent with the
22 kJ/mol obtained from temperature-dependent magnetic susceptibility measurements of 
the \textbf{bab} powder and with the 23 kJ/mol obtained previously from UCAM-B3LYP/6-31G(d,p)~\cite{bisbenzothia}. \textcolor{comm2}{This is also in line with the singlet $y_s$
always being considerably lower than the triplet values, suggesting that 
no singlet has bond length patterns corresponding to pure open-shell structures.}

The structural diradical characters $y_s$ for the B3LYP-D3 closed-shell solutions
match the experiment almost perfectly. For the longest molecule 
\textbf{bab}, the bs optimization converges to structures with 
a $\braket{\hat{S}^{2}}$ value of roughly 0.6, which is 
slightly lower in energy and features a larger $y_s$ than the
cs solution (0.71 vs.\ 0.63). Given the quite small energy differences,
this bs solution may be an artifact of DFT. It could also be that 
the good match of the cs data results from an error compensation between
the electronic structure description and the neglect of crystal packing
effects (compare the discussion in the preceding section). Without the experimental
data, there would be little solid criteria for 
deciding which of the two describes the experiment better.
As exact exchange admixtures increase in the functionals, the structural diradical characters 
of the closed-shell solutions increase slightly, leading to an overestimation of the
experimental values (that could still be consistent with the experiment if packing effects
should play a role). 

{$\mathrm{MAE}$ values between optimized and experimental geometries showed a very good correlation with $y_{el}$ values ranging from $1.00$ for \textbf{bbb} to $0.98$ for \textbf{bab}.
The correlation between $y_{s}$ and $y_{el}$ is slightly smaller ranging from $0.99$ for \textbf{bbb} to $0.88$ for \textbf{bab}.
This means that for organic systems, both $y_{s}$ and $\mathrm{MAE}$ are well suited.}

Bond length alternation decreases as the molecules get longer, which is 
consistent with the increasing diradical character. Also 
if BLA is taken as a criterion, closed-shell B3LYP-D3 matches the
experiment well, slightly overestimating BLA, while TPSSh errs in the
other direction showing the best agreement of the functionals considered.
For these organic systems, BLA is much more consistent over different
functionals, and the conclusions drawn from BLA and $y_s$ are similar:
the two hybrid functionals are suited best to describe bond length
patterns in these organic diradical candidates. This good agreement 
between the two measures may be because (1) BLA is more suitable for organic
systems than for inorganic ones, and because (2) the same sets of 
bonds are employed in evaluating these measures here, in contrast to the
diiron and dicobalt complexes discussed above. 
Furthermore, while experimental error bars on bond lengths still affect
BLA values more than $y_s$, this is much less severe than it was the
case for the two diiron complexes discussed above,
so for organic systems, both BLA and $y_s$ appear as reasonable 
choices for evaluating agreement between calculated and 
experimental bond length patterns. 

\FloatBarrier

\section{Conclusion}

For predicting nonlinear optical properties of molecules, 
it is essential to provide correct molecular structures
based on first-principles electronic structure methods.
For this purpose, small absolute errors
are not sufficient, but also a reliable description of relative 
structural parameters such as 
bond length
patterns is necessary.
%
We have therefore suggested a new measure,
 structural diradical character $y_s$, which is based
on comparisons between molecular structures and idealized 
closed-shell and diradical structures. 
We can show that with this new measure, consistent comparisons 
between experiment and first-principles molecular structures
for diradicals are possible. 

Based on these comparisons, we can identify 
two hybrid functionals, TPSSh and B3LYP, with 10 and 20 percent 
of exact exchange admixture, as suitable for describing 
structural diradical character in both organic and
organometallic systems. B3LYP \textcolor{comm2}{(with Grimme's empirical dispersion corrections)}
works best for 
the organic molecules, and TPSSh \textcolor{comm2}{(without dispersion correction)}
for the organometallic
complexes under study. 
\textcolor{comm2}{Importantly, 
these functionals were also the ones which gave a realistic 
description of singlet--triplet energy differences.}
The GGA and meta-GGA functionals BP86 and TPSS turned out
not suitable for \textcolor{comm2}{neither} purpose.

The excellent agreement for B3LYP\textcolor{comm2}{-D3}
was only found when the organic molecules (a series of
bisbenzothiaquinodimethanes with different molecular lengths)
were described as closed-shell electronic structures (restricted
KS-DFT), even though in some cases broken-symmetry 
solutions with partial open-shell singlet character
were slightly lower in energy, but within typical DFT error bars (up to 3.5 kJ/mol).
This illustrates
that present-day \ac{KS} \ac{DFT} may be unable to make predictions in cases
like these, where closed-shell and open-shell singlets are close in energy.
On the upside, there exists one frequently used functional, B3LYP, which
is able to provide perfect agreement for these structures when
only the closed-shell singlets are considered. Possibly, these data
could indicate that when singlet--triplet gaps are large, and when
closed- and open-shell singlets are nearly degenerate, one should
consider the closed-shell singlets as more reliable for present-day
standard \ac{xc} functionals. However,
such statements clearly require more research,
\textcolor{comm2}{possibly also considering schemes which combine
a more explicit description of static correlation with
Kohn--Sham DFT~\cite{li14,hube16,hede15,garz14,ston18}.}

\textcolor{comm2}{Comparing structural diradical character $y_s$ obtained from experiment}
with assignments as (predominantly) open-shell or 
closed-shell from the literature suggests that  $y_s$ smaller than roughly 
0.6 \textcolor{comm2}{(with MAE as a measure for structural
deviations)} points to 
a more closed-shell structure, while larger  $y_s$ correspond to 
more open-shell structures. \textcolor{comm2}{Closeness to triplet
$y_s$ values may also serve as an absolute criterion for pure open-shell character
(usually only applicable to computed structures, however).} 
For closed-shell electronic structures, 
$y_s$ slightly decreases with increasing exact exchange admixture, 
while for open-shell singlet, it increases, so that differences 
between the two structures become more pronounced.

Our work was motivated by our attempt at a true first-principles
prediction of the open-shell character for a photoswitchable [Co$_2$] complex,
which may be a structure worth pursuing and optimizing further
for achieving photoswitchable NLO properties. 
Our findings imply that 
B3LYP, which suggests an open-shell singlet ground state, is more
reliable than BP86, which favors the closed-shell singlet.
Therefore, further research into this and related compounds,
and their switchability, appears a worthwhile avenue of research. 

It is challenging to define diradical measures 
which are also applicable to experimental data (with a notable exception
suggested by Kamada et al.~\cite{yfromexp}).
Therefore, beyond such comparisons between theory and experiment, structural diradical character may also be interesting as a complement to electronic diradical character. This will require generalizing the definition of
reference structural parameters, e.g. for structures
where diradical character correlates with the presence or absence of a tin--tin bond rather than
bond length alternation.


\FloatBarrier

 \section{Acknowledgment}
The authors acknowledge funding by the German Research Foundation (DFG)
via SFB 668,
the high-performance-computing team of the
Regional Computer Center at University of Hamburg  and the
North-German Supercomputing Alliance (HLRN)
for technical support and computational resources.
The authors thank Markus Reiher, ETH Zurich, and J\"{u}rgen Heck,
University of Hamburg, for valuable discussions.

\begin{appendix}

\section{Critical discussion of 
bond length alternation (BLA) as a measure for 
comparing diradical structures}
\label{sec:bla}

\textcolor{comm3}{
\ac{BLA}~\cite{bladef1,bladef2} is evaluated as the average 
difference between bond lengths for $N$ pairs of adjacent bonds ($b_{i,0}$ and $b_{i,1}$),
\begin{equation}
\mathrm{BLA} = \frac{1}{N} \sum_{i=1}^{N} b_{i,0} - b_{i,1},
\label{eq:bla}
\end{equation}
which, e.g., can be alternating single and double or alternating single and triple bonds.\cite{bladef1,bladef2} The bonds that were considered for the \ac{BLA} are shown in the corresponding figures in red if they were added and in blue if they were subtracted. 
}

Using \ac{BLA} as a measure for diradical character has some drawbacks. (i) The \ac{BLA} can only be used for one sort of alternating bonds, while for the iron complexes (\textbf{[Fe\textsubscript{2}]} and \textbf{[Fe\textsubscript{2}]'}) there are both alternating single and double bonds, as well as alternating single and triple bonds. One has to choose which set of bonds will be used for calculating the \ac{BLA} and which will be dismissed. \textcolor{comm2}{Also, for some inorganic structures such as the tin cluster studied in Ref.~\cite{schr11}, diradical character correlates with the presence or absence of a tin--tin bond rather than
bond length alternation.} (ii) Pairs of bonds have to be used, meaning that an even number of bonds must be considered. For example, for the biscobaltocenyldithienylethene (\textbf{[Co\textsubscript{2}]}) this is not the case, and one has to choose arbitrarily one bond which will not be taken into account.
(iii) The sign of the calculated \ac{BLA} is choice-dependent and could be switched by adding the bond lengths that were subtracted and vice versa.
This indicates that it is difficult to define a unique and transferable measure for diradical character based on bond-length alternation.
(iv) The calculated numbers for different systems cannot be directly compared, because the magnitude of the numbers is system-dependent. \textcolor{comm2}{(v) \ac{BLA} measures bond length patterns in absolute terms, which implies that when comparing calculated with experimental data, deviations resulting from a general  over- or underestimation of bond lengths by a given functional are mixed with those resulting from an inadequate representation of
relative bond lengths differences.}

\section{Theoretical methods}

All electronic structure calculations were performed using the \textsc{Turbomole 6.5} program package \textcolor{comm}{imposing no symmetry ($C_1$ symmetry) on the systems. A value of $10^{-7}~\rm{a.u.}$ was set as the convergence criterion for the energy during self-consistent field calculations. For the molecular structure optimizations the threshold was set to $10^{-6}~\rm{a.u.}$ for the energy change, to $10^{-3}~\rm{a.u.}$ for the maximum displacement element, to $10^{-4}~\rm{a.u.}$ for the maximum gradient element, to $5 \cdot 10^{-4}~\rm{a.u.}$ for the root mean square of the displacement, and to $5 \cdot 10^{-4}~\rm{a.u.}$ for the root mean square of the gradient. The employed \ac{xc} functional (either BP86\cite{bp861,bp862}, TPSS\cite{tpss1}, TPSSh\cite{tpssh1,tpssh2} or B3LYP\cite{b3lyp1,b3lyp2,b3lyp3}) and basis set (either def-TZVP\cite{deftzvp1,deftzvp2} or def2-TZVP\cite{def2tzvp1,def2tzvp2}) and whether the third generation empirical dispersion correction of Grimme (D3)\cite{grimmed3} was used or not is indicated
in the respective tables.} 

For the closed-shell calculations restricted \ac{KS}-\ac{DFT} was employed. In order to obtain the broken-symmetry determinants, an unrestricted \ac{KS}-\ac{DFT} \textcolor{comm}{calculation} of the triplet state was performed and followed by a subsequent broken-symmetry calculation using Turbomole's ``flip'' option on the triplet determinant for obtaining the initial guess for the broken-symmetry calculation. The calculated ground state was then determined by comparison of the energies of the determinants and their corresponding $\braket{\hat{S}^{2}}$ values.  

\section{On the applicability of broken-symmetry density functional theory for
structural optimizations of diradicals}
\label{sec:bs}

A closed-shell determinant will always have a $\braket{\hat{S}^{2}}$ value of zero, 
while an open-shell determinant representing a diradicaloid will have 
a $\braket{\hat{S}^{2}}$ value larger than zero. 
This is referred to as spin contamination. For discussions on the
validity of broken-symmetry energies, see, e.g., 
Refs.~\cite{ruiz99,more06,perd95,jaco12, cohe07,wang95,baue83,goer93,davi05}.
\textcolor{comm2}{On the one hand, it is argued that the Kohn--Sham reference system
of noninteracting fermions should have the same $\braket{\hat{S}^{2}}$ value
as the real, interacting system,
and following this argument, schemes have been suggested for estimating the 
molecular structure of the spin-projected open-shell singlet based on 
broken--symmetry calculations~\cite{malr12}.}
 On the other hand, it is not generally 
established how to evaluate the $\braket{\hat{S}^{2}}$ value of the
interacting system in Kohn--Sham DFT, and there is no unique
established way for handling possible double counting
of electron correlation when employing spin projection 
on top of broken-symmetry Kohn--Sham determinants~\cite{li14,hube16,hede15,garz14}.
In practice, the Broken-Symmetry  approach has been very successful
in modeling molecular structures and energetics of
antiferromagnetically coupled systems~\cite{cram09,ruiz99}.
We therefore take a pragmatic approach here, directly evaluating the 
broken-symmetry energies as those of the open-shell singlet.
For future work, it would be interesting to consider
schemes which combine
a more explicit description of static correlation with
Kohn--Sham DFT~\cite{li14,hube16,hede15,garz14,ston18}.
At present, these are too computationally expensive for routine
structural optimizations of molecules of the size under study here.

\FloatBarrier

\end{appendix}


\begin{thebibliography}{100}
\bibitem{bla1}
For structures such as the tin cluster referred to above~\cite{schr11}, one
  could define reference bond lengths for typical tin--tin single bonds for the
  closed-shell and the sum of two tin van-der-Waals radii for the open-shell
  structure. This structure would clearly be a challenge for structural
  measures of diradical character.
\bibitem{n2}
Interpolated data between the two structures can be found in the Supporting
  Information.
\end{thebibliography}
\end{document}